\documentclass[twocolumn,pra,superscriptaddress,showpacs,letter,aps,10pt]{revtex4-2}

\usepackage{times}
\usepackage{graphicx}
\usepackage{dcolumn}
\usepackage{bm}
\usepackage{mathtools}
\usepackage[english]{babel}
\usepackage{physics}
\usepackage{enumitem}
\usepackage{mathrsfs}
\usepackage{xfrac}
\usepackage{xr-hyper}
\usepackage{hyperref}
\usepackage{footmisc}
\usepackage{hypcap}
\usepackage[dvipsnames]{xcolor}
\usepackage[normalem]{ulem}
\usepackage{comment}
\usepackage{amsmath}
\usepackage{siunitx}

\newcommand{\ci}{\mathrm{i}}

\newcommand{\tp}{\mathrm{p}}

\begin{document}

\title{Momentum correlations of the Hawking effect in a quantum fluid}

\author{Marcos Gil de Olivera}
\affiliation{Laboratoire Kastler Brossel, Sorbonne Universit\'{e}, CNRS, ENS-Universit\'{e} PSL, Coll\`{e}ge de France, Paris 75005, France}
\affiliation{Instituto de F\'{\i}sica, Universidade Federal Fluminense, 24210--346 Niter\'{o}i, RJ, Brazil}

\author{Malo Joly}
\affiliation{Laboratoire Kastler Brossel, Sorbonne Universit\'{e}, CNRS, ENS-Universit\'{e} PSL, Coll\`{e}ge de France, Paris 75005, France}

\author{Antonio Z. Khoury}
\affiliation{Instituto de F\'{\i}sica, Universidade Federal Fluminense, 24210--346 Niter\'{o}i, RJ, Brazil}

\author{Alberto Bramati}
\affiliation{Laboratoire Kastler Brossel, Sorbonne Universit\'{e}, CNRS, ENS-Universit\'{e} PSL, Coll\`{e}ge de France, Paris 75005, France}

\author{Maxime J. Jacquet}
\email[]{maximjacquet@gmail.com}
\affiliation{Laboratoire Kastler Brossel, Sorbonne Universit\'{e}, CNRS, ENS-Universit\'{e} PSL, Coll\`{e}ge de France, Paris 75005, France}

\date{\today}

\begin{abstract}
The Hawking effect -- the amplification of fluctuations at the horizon -- has been detected in quantum fluids through real-space density correlations.
However, real-space observables integrate over frequency, mixing distinct scattering channels into a single interference pattern and obscuring the spectral and entanglement structure of the emission.
Here, we numerically compute momentum-space two-point correlations in a transcritical quantum fluid, using the truncated Wigner approximation applied to a realistic, driven-dissipative polariton system.
We spectrally resolve the Hawking--partner as well as greybody factor channels and find that both carry comparable correlation strength, demonstrating that the well-known real-space ``moustache'' is an interference between these spectrally distinct contributions.
The Hawking--partner channel is dominated by negative correlations, a direct signature of quantum-vacuum pair creation.
All features reach amplitudes detectable in state-of-the-art experiments.
Our results establish that the full three-mode output state must be considered for entanglement characterization, and provide a general framework applicable to any quantum fluid.
\end{abstract}

\maketitle

The Hawking effect -- the emission of correlated pairs of field excitations from horizons -- is a cornerstone prediction of quantum field theory in curved spacetime~\cite{Hawking,wald_particle_1975,WaldBook}.
In quantum fluids of light or matter, an acoustic horizon forms when the flow transitions from subcritical to supercritical~\cite{Unruh,visser_acoustic_1998}, and the Hawking effect remains robust in dispersive media~\cite{brout_hawking_1995,unruh_sonic_1995,corley_hawking_1996,barcelo_analogue_2011}.
In such media with superluminal dispersion, scattering at the horizon produces three outgoing modes at each frequency: Hawking radiation (HR) propagating upstream, its negative-energy partner, and a positive-energy `witness' mode~\cite{Recati_2009} associated with  greybody factors~\cite{agullo_event_2022}.

To date, the experimental detection of the Hawking effect in quantum fluids has relied on real-space density-density correlations~\cite{munoz_de_nova_observation_2019,isoard_departing_2020}, which produce a so-called ``moustache'' interference pattern~\cite{carusotto_numerical_2008,balbinot_nonlocal_2008}.
Meanwhile, the presence of three outgoing modes -- and the importance of the witness channel in particular -- has been established in analytical studies of closed atomic BECs~\cite{Recati_2009,larre_quantum_2012,michel_phonon_2016,Isoard}, and the resulting tripartite entanglement structure has been studied extensively~\cite{isoard_bipartite_2021,agullo_event_2022}.
However, real-space observables integrate over all frequencies: correlations between the HR, partner and witness modes (which share the same frequency but propagate at different group velocities on either side of the horizon) overlap spatially and cannot be separated~\cite{balbinot_nonlocal_2008,isoard_departing_2020,isoard_bipartite_2021}.
The moustache is therefore not a clean image of any single correlation channel but an interference pattern between spectrally distinct contributions~\cite{isoard_departing_2020,isoard_bipartite_2021,jacquet_quantum_2023}.
Isolating these channels and assessing their relative weight requires momentum-space analysis.

Analytical predictions for momentum correlations have been developed for conservative systems with idealized profiles: analytic horizon shapes with homogeneous asymptotic regions~\cite{boiron_quantum_2015,fabbri_momentum_2018}.
These treatments resolve the cross-horizon correlations -- the dominant anomalous Hawking--partner channel, an entanglement witness~\cite{Busch_entanglementHR_2014}, as well as the HR--witness channel~\cite{fabbri_momentum_2018} -- but do not consider correlations between modes propagating on the same side of the horizon.
Moreover, no real experiment satisfies their idealizations: laboratory systems are finite, horizons have spatial structure~\cite{finazzi_spectral_2011,jacquet_quantum_2023}, and photonic platforms are intrinsically driven-dissipative~\cite{jacquet_superradiant_2025}.

Here, we provide a numerical computation of the full momentum-space correlation map of a transcritical quantum fluid.
We use the truncated Wigner approximation (TWA) applied to a driven-dissipative polariton system, in a configuration that has been experimentally realized in~\cite{falque_polariton_2025}.
We obtain spectrally resolved correlation maps with three key findings:
(i)~the Hawking-partner and Hawking-witness (greybody factor) correlation traces are both clearly resolved and carry comparable correlation amplitudes -- the witness mode cannot be neglected;
(ii)~the Hawking-partner trace is dominated by negative correlations, indicating quantum-vacuum pair creation;
(iii)~all features are at the level $|\delta g_2|\sim 10^{-3}$, which is detectable in state-of-the-art experiments.

\textit{Acoustic horizon in a polariton fluid.}---
Polaritons -- hybrid light-matter quasiparticles arising from the strong coupling of cavity photons and quantum-well excitons~\cite{carusotto_quantum_2013} -- offer distinct advantages for analog gravity studies~\cite{jacquet_superradiant_2025}.
Their small effective mass ($m^*\approx 5\times 10^{-5}\,m_e$) yields healing lengths $\xi\sim\SI{1}{\micro\meter}$, so the horizon structure extends over a few micrometers and is fully resolved optically.
Crucially, the photonic component of polaritons decays at rate $\gamma$, continuously emitting photons that carry the full quantum state of the fluid into the far field.
This provides direct optical access to both density and phase -- and hence to real- and momentum-space correlations -- without destructive measurements, in contrast to time-of-flight techniques used in atomic BECs~\cite{falque_polariton_2025}.

We consider a one-dimensional polariton fluid in a semiconductor microcavity, coherently driven by a continuous-wave laser.
The mean field $\psi$ obeys a dynamic dictated by a driven-dissipative Gross-Pitaevskii equation  (GPE)~\cite{carusotto_quantum_2013} with repulsive interactions proportional to the interaction energy $\hbar gn$ ($n(x,t)=\abs{\psi(x,t)}^2$ the local density, $g>0$), photonic losses $\gamma$ and external driving $F(x,t)$
\begin{align}
    \label{eq:ddgpe}
    \ci \partial_t \psi(x, t) = &\Big( \omega_0 - \frac{\hbar \partial_x^2}{2m^*}+ V(x) + g n(x,t)  - \frac{\ci\gamma}{2} \Big) \psi(x,t)\nonumber\\+F(x,t),
\end{align}
with $\omega_0$ the frequency of the bottom of the polariton branch.

\begin{figure*}[ht!]
    \centering
    \includegraphics[width=0.9\linewidth]{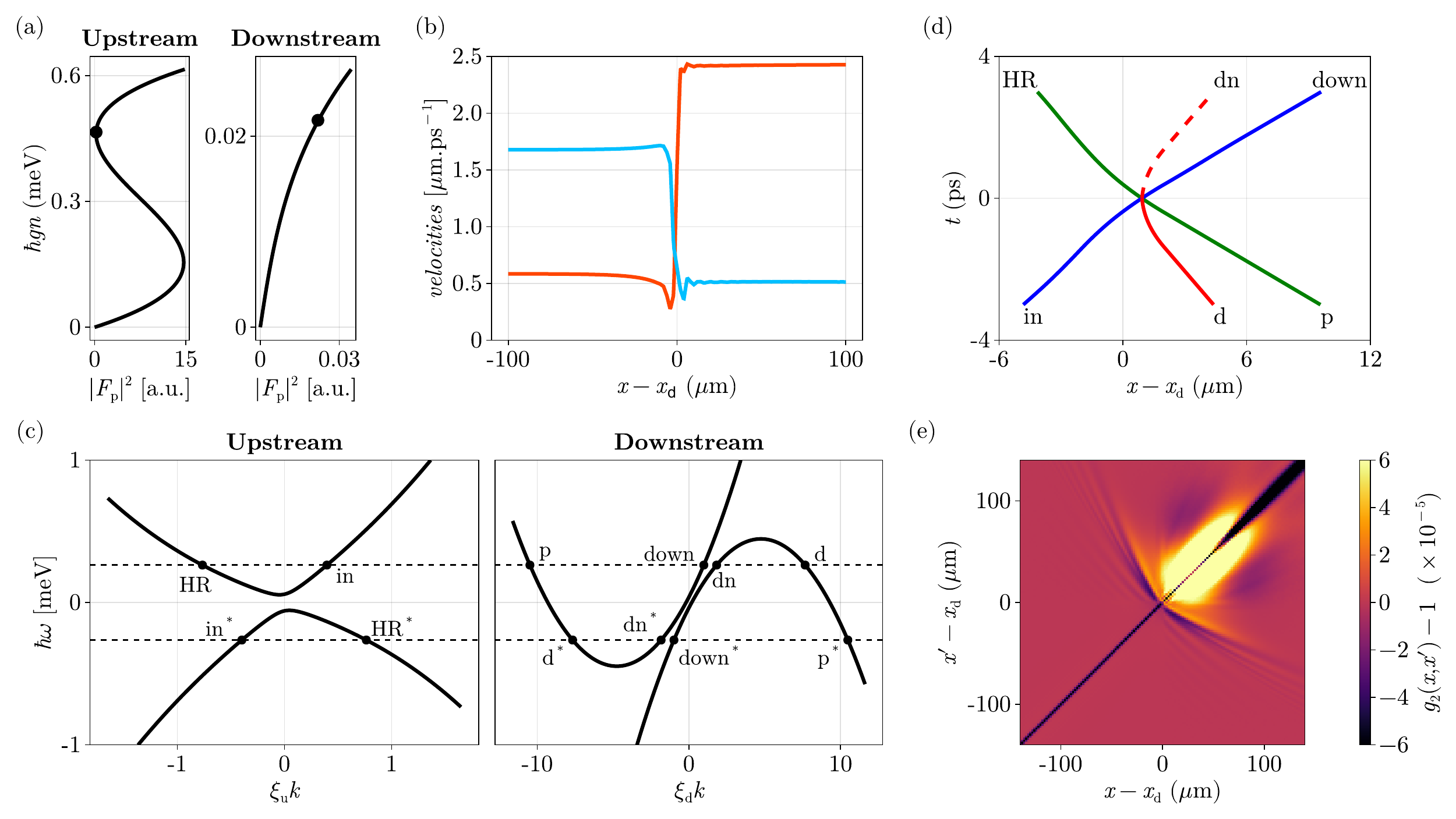}
    \caption{%
    \textbf{Acoustic horizon and real-space correlations.}
    (a)~Flow velocity $v$ (red) and Bogoliubov sound speed $c_\mathrm{B}$ (blue) vs.\ position relative to the defect $x_\mathrm{d}$; the crossing defines the acoustic horizon.
    (b)~Dispersion relation upstream (left) and downstream (right); dashed lines show a fixed $\omega$.
    (c)~WKB trajectories at $\omega=\SI{0.4}{\per\pico\second}$: the incoming mode (blue) scatters into Hawking radiation HR (green), its negative-norm partner dn (red, dashed), and a positive-energy witness mode down (blue, right).
    (d)~Real-space $g_2(x,x')-1$: the ``moustache'' in the upper-left and lower-right quadrants is an interference pattern between HR--dn and HR--down correlations at all frequencies.}
    \label{fig:big_picture}
\end{figure*}

As shown in Fig.~\ref{fig:big_picture}(a), an inhomogeneous pump with structured wavevector $k_\tp(x)$ and intensity $|F(x)|^2$, together with a localized repulsive potential $V(x)$ (height \SI{0.85}{\milli\electronvolt}, width \SI{0.75}{\micro\meter}), creates a stationary profile comparable to the experiment~\cite{falque_polariton_2025} and reproducing the numerics of~\cite{jacquet_analogue_2022}, where the flow velocity \(v(x) = \hbar \, \partial_x (k_\tp(x)\,x)/m^*\) (red) crosses the sound cone $c_\mathrm{B}(x)$ (blue)~\footnote{In each spatial region, the pump frequency $\omega_\tp(x)=cst$ defines an effective detuning $\delta(k_\tp) = \omega_\tp - \omega_0 - \hbar k_\tp(x)^2/(2m^*)$ that controls the sound cones and the healing length via $c_\mathrm{B}=\sqrt{\hbar (2gn-\delta(k_\tp))/m^*}$ and $\xi=\sqrt{\hbar/m^*(2gn-\delta(k_\tp))}$, respectively, see~\cite{falque_polariton_2025} and SM.}. On the left hand side, $v<c_\mathrm{B}$ so the upstream region is subcritical, while on the right hand side $v>c_\mathrm{B}$ so the downstream region is supercritical.

The stationary solution of the mean field equation serves as a background on top of which perturbations propagate. These collective excitations experience an effective curved spacetime described by the Painlev\'e--Gullstrand metric
$ds^2=(c_\mathrm{B}^2-v^2)\,dt^2 - 2v\,dx\,dt + dx^2$~\cite{visser_acoustic_1998,falque_polariton_2025,jacquet_superradiant_2025}\footnote{The polariton lifetime is $\tau\approx\SI{15}{\pico\second}$.
Correlations are sampled over $\lesssim 3\tau$, during which the continuous-wave pump locks the mean field.
The weak Hawking flux produces no measurable back-reaction.}. The crossing from the subcritical to the supercritical region then defines a Killing horizon for the excitations.

In this transcritical flow, the laboratory-frame dispersion [Fig.~\ref{fig:big_picture}(b)]
\begin{equation}
    \label{eq:lfbogo}
        \omega^{\pm}(q)=v(x) q\pm\sqrt{\left[\frac{\hbar q^2}{2m^*}-\delta(k_\mathrm{p}) +2gn(x)\right]^2-[gn(x)]^2}
\end{equation}
is Doppler shifted. Here, $q$ denotes the momentum in the fluid frame, which relates to the momentum in the laboratory frame through the expression $k = k_\tp + q$.
Modes on $\omega^+$ have positive symplectic norm~\cite{castin_lecture_notes} $Q = \int dx\,(|u_q|^2 - |v_q|^2)$; those on $\omega^-$ have negative norm~\cite{jacquet_superradiant_2025}.
Notably, in the supercritical region, the negative-norm Bogoliubov branch is drawn to positive laboratory-frame frequencies, up to $\omega_\mathrm{max}$, allowing negative-energy modes to be excited, while positive-energy modes can be excited upstream for $\omega\geq\omega_\mathrm{min}$ ($\omega_\mathrm{min}\geq0$ is set by the minimum of the upstream positive-norm branch, i.e. the effective mass of the Bogoliubov field)~\cite{falque_polariton_2025}.

The shape of the dispersion on either side of the horizon defines the interval $[\omega_\mathrm{min},\omega_\mathrm{max}]$ inside which two modes (\textbf{HR} and \textbf{in}) propagate upstream and four downstream (\textbf{p} and \textbf{down}, \textbf{d} and \textbf{dn}), kinematically allowing the norm-mixing scattering responsible for the Hawking effect~\cite{jacquet_analogue_2022,jacquet_superradiant_2025}.
Outside this interval, norm-mixing across the horizon is not possible and spontaneous pair creation is suppressed.
The width of this window -- set by the Mach number and the effective detuning -- determines the spectral extent of the correlation traces in momentum space.
Differently from massless field theories, the spectrum peaks near $\omega_\mathrm{min}$.

\textit{Three-mode scattering at the horizon.}---
The Hawking effect can be understood from the three-channel scattering picture in Fig.~\ref{fig:big_picture}(c): the Hawking progenitors \textbf{p} (positive-norm) and \textbf{d} (negative-norm) impinge on the horizon from the supercritical region (as in all systems with superluminal dispersion~\cite{brout_hawking_1995,corley_hawking_1996,porrotunneling2024}), get squeezed by the horizon~\cite{jacquet_superradiant_2025,agullo_entanglement_2023} and mix with the positive-norm mode \textbf{in} coming from the subcritical region.
This norm-mixing scattering process yields amplification in the outgoing modes \textbf{HR}, \textbf{down} and \textbf{dn}.

At each frequency $\omega\in[\omega_\mathrm{min},\omega_\mathrm{max}]$, the scattering at the horizon is captured by a $3\times3$ matrix~\cite{Recati_2009}
\begin{equation}\label{eq:SM}
    \begin{pmatrix}
        \hat b_{\text{HR}}\\
        \hat b_{\text{down}}\\
        \hat b^{\dagger}_{\text{dn}}
    \end{pmatrix}
    = S_\omega
    \begin{pmatrix}
        \hat a_{\text{in}}\\
        \hat a_{\text{p}}\\
        \hat a^{\dagger}_{\text{d}}
    \end{pmatrix},
\end{equation}
connecting three incoming modes to three outgoing modes.
The appearance of $\hat{b}^\dagger_\mathrm{dn}$ and $\hat{a}^\dagger_\mathrm{d}$ reflects anomalous Bogoliubov mixing responsible for spontaneous pair creation.
The dominant quantum correlation is anomalous: $\langle \hat{b}_\mathrm{HR}\hat{b}_\mathrm{dn}\rangle\neq 0$, which violates the Cauchy--Schwarz inequality and witnesses the nonseparability of the output state~\cite{boiron_quantum_2015,fabbri_momentum_2018,Busch_entanglementHR_2014}.

Crucially, $S_\omega$ is a three-mode matrix.
\textbf{HR} is also classically correlated with the positive-norm witness mode \textbf{down}~\cite{isoard_bipartite_2021}.
In real space, the HR--dn$^*$ and HR--down correlations at the same frequency $\omega$ both appear in the same quadrant of the correlation map, but over different cones set by the group velocity (dispersion relation) of each involved mode.
The corresponding correlation traces therefore follow different curves in the $(x,x')$ plane, but these curves overlap and interfere, producing the moustache pattern visible in the upper-left and lower-right quadrants of Fig.~\ref{fig:big_picture}(d).
What has been interpreted as the Hawking--partner signal is, in fact, this interference~\cite{isoard_departing_2020,isoard_bipartite_2021,jacquet_quantum_2023}.

For the present configuration, as in any realistic implementation, the mean-field profile is highly inhomogeneous, so Eq.~\eqref{eq:SM} cannot be computed analytically.
We simulate the full quantum dynamics using the TWA [SM], a stochastic method valid for weak interactions and large occupation numbers that naturally handles dissipation, finite-size effects, and arbitrary profiles~\cite{carusotto_quantum_2013}.
The code is available at~\cite{Gil_de_Oliveira_momentum_correlation_polaritons_2025, Gil_de_Oliveira_GeneralizedGrossPitaevskii_jl_2025}.

\begin{figure}
    \centering
    \includegraphics[width=\columnwidth]{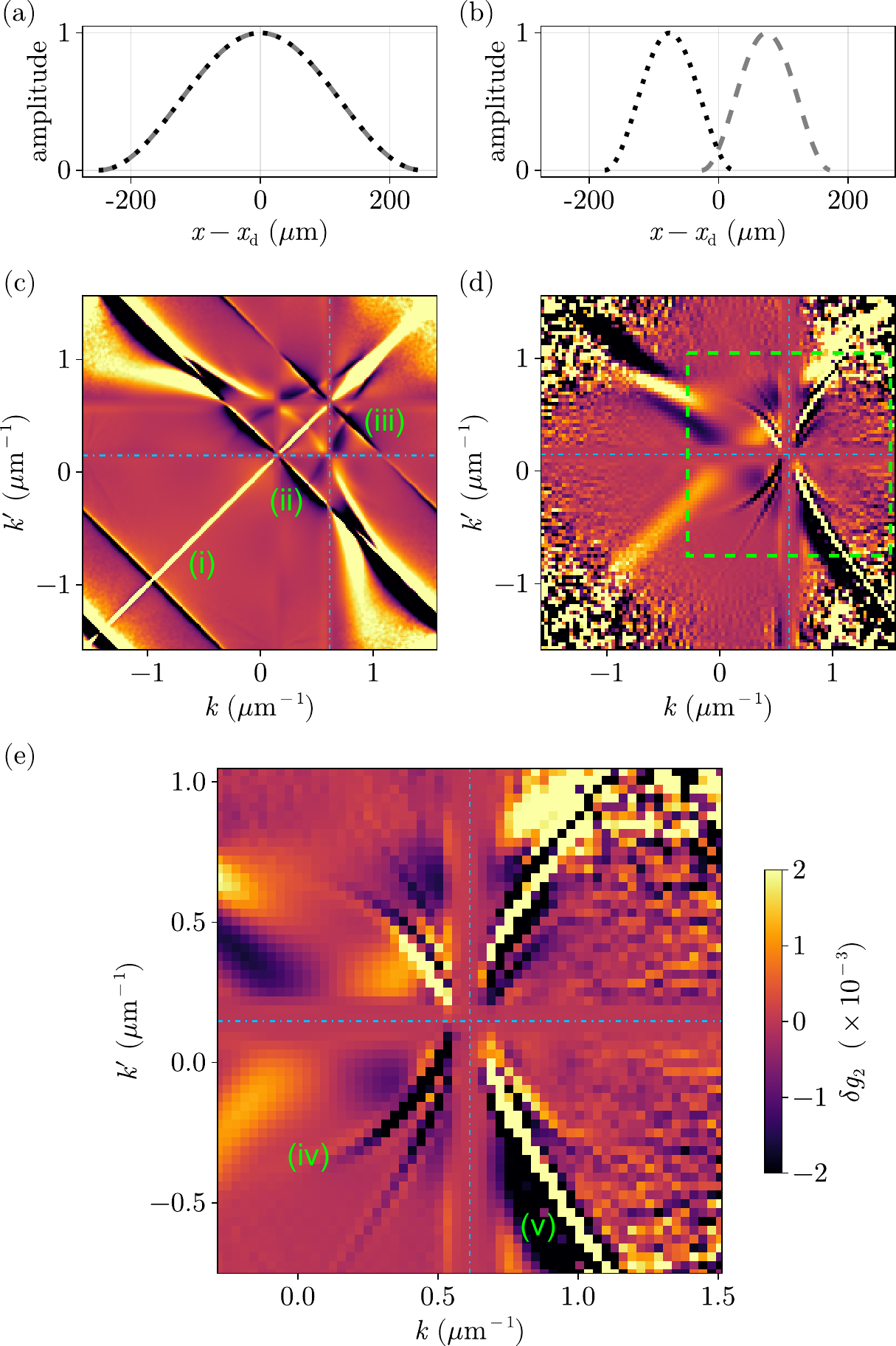}
    \caption{%
    \textbf{Momentum-space Hawking correlations.}
    (a,b)~Spatial window functions.
    (c)~$\delta g_2(k,k')$ with overlapping windows: features (i)--(iii) are dominated by autocorrelations.
    (d)~With separated windows, autocorrelations are suppressed.
    (e)~Zoom on the cross-horizon region.
    Traces~(iv): \textbf{HR}--\textbf{dn}$^*$ (Hawking--partner), dominated by negative correlations (quantum-vacuum signature).
    Traces~(v): \textbf{HR}--\textbf{down} (Hawking--witness), at comparable amplitude.
    Dot-dashed lines: $k_\mathrm{up}$, $k_\mathrm{down}$.}
    \label{fig:large_windows}
\end{figure}

\textit{Spectrally resolved correlations.}---We compute the momentum-space density correlation
\begin{equation}
\delta g_2 \equiv g_2(k,k')-1=
\frac{\left\langle \hat\psi_\mathrm{U}^\dagger(k')\,\hat\psi_\mathrm{D}^\dagger(k)\,\hat\psi_\mathrm{D}(k)\,\hat\psi_\mathrm{U}(k') \right\rangle}
{\left\langle \hat\psi_\mathrm{D}^\dagger(k)\hat\psi_\mathrm{D}(k)\right\rangle\left\langle \hat\psi_\mathrm{U}^\dagger(k')\hat\psi_\mathrm{U}(k')\right\rangle}-1,
\label{eq:delta_g2}
\end{equation}
where $\hat\psi_{\mathrm{U},\mathrm{D}}(k)$ are windowed Fourier transforms of the field on the upstream (U) and downstream (D) sides of the horizon [SM].
Fig.~\ref{fig:large_windows} shows $\delta g_2$ for two spatial (Hann) window configurations: equal windows centered on the horizon [panel~(a)$\to$(c)], and windows placed on opposite sides [panel~(b)$\to$(d,e)].

With overlapping windows [Fig.~\ref{fig:large_windows}(c)], the map displays symmetric features (i)--(iii) dominated by autocorrelations.

Feature~(i) is the diagonal ($k=k'$) autocorrelation of each spatial region with itself, broadened by the finite window size into the bright ridge along the main diagonal.
Feature~(ii) is the line $(k-k_\mathrm{U}) = -(k^\prime-k_\mathrm{U})$ while~(iii)  is the line $(k-k_\mathrm{D}) = -(k^\prime-k_\mathrm{D})$.
They both arise from four-wave mixing on the same side of the horizon~\cite{jacquet_analogue_2022,claude2021highresolution}: two photons with momentum $k_\tp$ produce one photon with momentum $k = k_\tp + q$ and another one with momentum $k^\prime = k_\tp - q$.
These autocorrelation features, while unrelated to the Hawking effect, serve as internal benchmarks: their momenta and amplitudes are determined by the mean-field profile and occupation numbers, providing a consistency check on the simulation.
They are not innocuous, however: near the pump wavevectors, where the correlation tongues converge, traces~(ii) and~(iii) may overlap the HR--dn$^*$ and HR--down traces, and with overlapping windows their contribution mixes with the cross-horizon signal.
Hawking correlations must therefore be evaluated away from these crossings, or with window configurations that suppress the same-side contributions [Fig.~\ref{fig:large_windows}(d)].

With separated windows [Fig.~\ref{fig:large_windows}(d)], the cross-convolution of identical windows cancels self-overlap terms, suppresses autocorrelations, and allows the cross-horizon correlations to emerge.
Increasing the separation between windows and the horizon decreases the correlations and increases the noise, reflecting the finite polariton lifetime, a feature absent in closed systems~\cite{fabbri_momentum_2018} [SM].

The zoomed panel (e) reveals our central result: two spectrally resolved families of correlation with Hawking radiation.

\textit{Two channels of comparable strength.}---Traces~(iv) correspond to the anomalous HR--dn$^*$ (Hawking--partner) channel, while traces~(v) correspond to HR--down (greybody factors).
Both carry amplitudes $|\delta g_2|\sim 10^{-3}$.
Each physical trace appears in four quadrants related by the sign-reversal symmetry $(q,q')\mapsto(\pm q,\pm q')$, yielding eight traces total -- a consistency check enforced by the bosonic commutation relations~\cite{fabbri_momentum_2018,macher_black/white_2009}.

Quantitatively, the HR--dn$^*$ traces~(iv) reach peak amplitudes $|\delta g_2|\approx 2\times 10^{-3}$, read off along the portions of the traces free from same-side contributions, with the negative (anticorrelation) lobe being slightly stronger than the positive one.
The HR--down traces~(v) have peak values of the same order, $|\delta g_2|\approx 1.5\times 10^{-3}$, but are predominantly positive, consistent with the normal (classical) character of this correlation channel~\cite{isoard_bipartite_2021}.
Averaging $|\delta g_2|$ along each trace confirms this comparison: $\langle|\delta g_2|\rangle\approx 1.2\times 10^{-3}$ for HR--dn$^*$ and $\approx 8\times 10^{-3}$ for HR--down, so the two channels carry weight within one order of magnitude of one another along their full spectral extent and comparable weight at their maxima.
However, note that the HR--down tongue runs close to the same-side feature~(ii), so the measured value might be contaminated, as discussed above.
The momentum loci of the traces agree with the dispersion relation~\eqref{eq:lfbogo}: trace~(iv) connects $k_\mathrm{HR}(\omega)$ on the upstream positive-norm branch to $-k_\mathrm{dn}(\omega)$ on the downstream negative-norm branch -- the sign reversal reflecting the anomalous pairing -- while trace~(v) connects $k_\mathrm{HR}(\omega)$ to $k_\mathrm{down}(\omega)$ on the downstream positive-norm branch, both swept over the frequency window $[\omega_\mathrm{min},\omega_\mathrm{max}]$.

That both channels are visible with comparable strength is significant.
The HR--down correlations have been studied analytically in real-space configurations~\cite{Recati_2009,larre_quantum_2012,michel_phonon_2016,Isoard}, and the tripartite entanglement structure they imply is well established theoretically~\cite{isoard_bipartite_2021,agullo_event_2022}.
In momentum space, they have been computed analytically for idealized transcritical profiles~\cite{fabbri_momentum_2018}.
Our simulations resolve them in a realistic, driven-dissipative flow and show that they are stronger than the HR--dn$^*$ signal, which confirms numerically what the tripartite entanglement literature anticipated: the witness mode (greybody factor) is not a subdominant correction.
Neglecting it leads to an incomplete picture of the quantum state produced at the horizon.

The comparable strength of both channels might also explain why they have been difficult to separate experimentally~\cite{isoard_departing_2020}, as the correlation structure revealed here is universal to systems with superluminal dispersion.
In real-space measurements, the moustache pattern integrates both contributions at every frequency.
If one channel dominated overwhelmingly, the other could be treated as a perturbation.
Instead, the interference between two channels of similar amplitude produces the full moustache: its shape, sign structure, real-space amplitude, and position encode the combined HR--dn$^*$ and HR--down correlations, not either one in isolation.

\textit{Quantum-vacuum signature and horizon structure.}---The HR--dn$^*$ traces~(iv) are dominated by negative correlations, the hallmark of anomalous pair correlations $\langle \hat{b}_\mathrm{HR}\hat{b}_\mathrm{dn}\rangle$ arising from quantum vacuum amplification~\cite{fabbri_momentum_2018}.
Some traces also exhibit mixed correlation signs along their length --coexisting positive and negative regions characteristic of a Fano-like interferometric response.
This mixed-sign structure arises from the interplay between the finite-width horizon and Bogoliubov scattering, and is absent in analytical treatments based on analytic profiles approximating realistic implementations~\cite{fabbri_momentum_2018,isoard_departing_2020}, where the horizon has no internal structure.
It demonstrates that the near-horizon geometry leaves a spectrally resolved imprint on the Hawking correlations, connecting to recent work on quasi-normal modes~\cite{jacquet_quantum_2023,burkhard2025stimulatedhawkingeffectquasinormal} and broadened horizons~\cite{finazzi_spectral_2011,porrotunneling2024}.

This contrasts with the momentum-space predictions in atomic BECs, which assume analytical horizon profiles with uniform asymptotic densities~\cite{boiron_quantum_2015,fabbri_momentum_2018}.
In those idealized settings, the HR--dn$^*$ correlation follows a smooth, monotonic curve entirely determined by the analytical $S$-matrix elements.
The richer substructure we observe---sign changes, asymmetric lobes, and frequency-dependent modulation along the correlation traces---is a direct consequence of the finite spatial extent of the horizon and the inhomogeneous mean-field profile, features that are generic to all experimental implementations.
Our results suggest that a momentum-space analysis of experimental data would similarly reveal the two-channel structure and the spectral imprint of the horizon geometry.

This two-channel structure is not specific to the present parameter set.
The weight of the HR--down channel is set by the greybody factor of the horizon~\cite{Recati_2009,agullo_event_2022}, which is of order unity whenever the horizon region is not adiabatically smooth: steeper horizons increase the overall correlation amplitude, but both channels grow together and remain of comparable strength~\cite{fabbri_momentum_2018,jacquet_analogue_2022}.
Only in the limit of a smooth, adiabatic crossover, where norm-mixing scattering is suppressed, would the witness channel -- and with it the interference structure of the moustache -- fade.

\textit{Experimental detectability.}---All significant features satisfy $|\delta g_2|\geq 10^{-3}$.
In the laboratory, this is accessed by selecting two far-field momentum windows and measuring the zero-delay cross-covariance of the photocurrents in a balanced receiver.
The signal-to-noise scales as $\mathrm{SNR}\propto \delta g_2 \sqrt{R_u R_d\,\tau_c\,T}$, where $R_{u,d}$ are detected photon rates, $\tau_c$ the correlation time, and $T$ the integration time [SM].
For photon fluxes $\sim 10^{13}\,\mathrm{s}^{-1}$ per window ($\sim\SI{10}{\micro\watt}$ at $\sim\SI{830}{\nano\meter}$~\cite{falque_polariton_2025}), a $5\sigma$ detection requires $T\sim$ minutes [SM].

\textit{Discussion.}---Our results carry four implications that go beyond the specific system studied.

First, they provide direct evidence that the real-space moustache pattern (the standard observable for Hawking correlations) is an interference between the HR--partner and HR--witness (greybody factors) channels.
This is visible only because momentum-space analysis separates contributions that are inextricably mixed in position space.
The finding reframes how real-space measurements should be interpreted: the moustache encodes the coherent sum of at least two scattering channels at each frequency, not the Hawking--partner correlation alone.
This confirms previous analysis stating that artificial cancellation of this correlation contribution can lead to a misinterpretation of the spectrum of emission~\cite{isoard_departing_2020}.

Second, the comparable strength of the HR--down channel confirms numerically that the Hawking output state is genuinely three-mode.
This corroborates recent theoretical predictions on tripartite entanglement~\cite{isoard_bipartite_2021,agullo_event_2022}, and extends them to the experimentally relevant driven-dissipative regime~\cite{falque_polariton_2025,jacquet_superradiant_2025}.
Entanglement quantification based solely on the HR--dn pair misses the witness mode's contribution.
The spectrally resolved correlations computed here provide direct access to the full three-mode covariance matrix needed for complete entanglement characterization.

Third, the driven-dissipative nature of the polariton platform, rather than being merely a complication, plays a clarifying role.
Dissipation attenuates correlations exponentially with the propagation distance from the horizon, imposing a practical constraint: the spatial windows must be placed within a few decay lengths ($v_g\,\tau\sim 10$--$\SI{30}{\micro\meter}$, depending on the group velocity $v_g$ of each mode) of the horizon to maintain a detectable signal.
Conversely, dissipation suppresses recurrence effects and reflections from boundaries that complicate the interpretation of correlations in finite-size conservative systems~\cite{michel_phonon_2016,Isoard}.
The continuous-wave pump also prevents mean-field depletion, making the stationary-state assumption exact, and ensuring that the sampled fluctuations are purely quantum noise on a well-defined background.
Experimentally, full-field measurements (spatially and spectrally resolved non-commuting observables) are routinely implemented in various effective geometries for the quantum field~\cite{falque_polariton_2025,guerrero_multiply_2025}.

Finally, the TWA approach is general: it applies to any quantum fluid in the regime of weak interactions and large occupation numbers, whether conservative or driven-dissipative.
It naturally incorporates finite-size effects, realistic horizon profiles, and dissipation -- features that analytical treatments on homogeneous, infinite systems cannot capture~\cite{boiron_quantum_2015,fabbri_momentum_2018,macher_black/white_2009,Busch_entanglementHR_2014,nova_entanglement_2015}.
The method extends to rotating two-dimensional geometries~\cite{svancara_rotating_2024,guerrero_multiply_2025}, opening the way to frequency-resolved studies of superradiant amplification with and without horizons~\cite{delhom_entanglement_2024,agullo_entanglement_2023}.

\section*{Acknowledgements}

We are thankful to K\'evin Falque, Killian Guerrero, Iacopo Carusotto, and Nicolas Pavloff for their insights on polaritons and the Hawking effect.
The authors also thank Killian Guerrero for feedback on the manuscript.
MJJ and AB acknowledge funding from the EU Pathfinder 101115575 Q-One.
AB acknowledges support from the Institut Universitaire de France. MGO acknowledges the sponsorship by Coordena\c{c}\~ao de Aperfei\c{c}oamento de Pessoal do Ensino Superior - CAPES of his six-month internship at Laboratoire Kastler Brossel - LKB. AZK acknowledges support from Conselho Nacional de Desenvolvimento Cientifico e Tecnol\'ogico (CNPq 422300/20217); Funda\c{c}\~ao Carlos Chagas Filho de Amparo \`a Pesquisa do Estado do Rio de Janeiro (FAPERJ).

\end{document}


\title{Supplemental Material}
\author{Marcos Gil de Olivera}
\affiliation{Instituto de F\'{i}sica, Universidade Federal Fluminense, 24210-346 Niter\'{o}i, RJ, Brazil}
\affiliation{Laboratoire Kastler Brossel, Sorbonne Universit\'{e}, CNRS, ENS-Universit\'{e} PSL, Coll\`{e}ge de France, Paris 75005, France}

\author{Malo Joly}
\affiliation{Laboratoire Kastler Brossel, Sorbonne Universit\'{e}, CNRS, ENS-Universit\'{e} PSL, Coll\`{e}ge de France, Paris 75005, France}

\author{Antonio Z. Khoury}
\affiliation{Instituto de F\'{i}sica, Universidade Federal Fluminense, 24210-346 Niter\'{o}i, RJ, Brazil}

\author{Alberto Bramati}

\author{Maxime J. Jacquet}
\email[]{maxime.jacquet@lkb.upmc.fr}
\affiliation{Laboratoire Kastler Brossel, Sorbonne Universit\'{e}, CNRS, ENS-Universit\'{e} PSL, Coll\`{e}ge de France, Paris 75005, France}

\date{\today}
\maketitle

\section{Polariton Kinematics}

Polaritons are quasiparticles  of mass $m^*$ resulting from the strong coupling of cavity photons and quantum well excitons. Under coherent excitation by a pump $F(x, t)$ near resonance with the cavity, the mean field of polaritons can be described by the driven dissipative Gross-Pitaevskii equation~\cite{carusotto_quantum_2013}
\begin{equation}
    \label{eq::ddgpe}
    i \partial_t \psi(x, t) = F(x,t) + \left( \omega_0 - \frac{i\gamma}{2} - \frac{\hbar \partial_x^2}{2m^*} + g \abs{\psi(x,t)}^2 + V(x) \right) \psi(x,t).
\end{equation}
$\omega_0$ is the polariton frequency at wavenumber $k=0$, $\gamma$ the (mostly photonic) decay rate, $g > 0$ the repulsive interaction strength and $V(x)$ the potential reigning in the cavity.

We consider a monochromatic pump $F(x, t) = F(x)e^{-i\omega_p t}$. Then, it is useful to decompose the polariton field as $\psi(x,t) = \sqrt{n(x,t)} e^{i(\theta(x,t) - \omega_p t)}$. The density $n$ and the phase $\theta$ satisfy the equations
\begin{equation}
    \label{eq:continuity}
    \partial_t n + \partial_x  (n v) = -\gamma n + 2\sqrt{n}\Im Fe^{-i\theta} ;
\end{equation}
\begin{equation}
    \label{eq:euler}
    \partial_t \theta + gn - \delta_0 + \frac{m^*v^2}{2 \hbar} + V - \frac{\hbar}{2m^*}\frac{\partial_x^2 \sqrt{n}}{\sqrt{n}} +  \frac{\Re F e^{-i\theta}}{\sqrt{n}} = 0.
\end{equation}
We have defined the velocity of the fluid as $v = \hbar \partial_x \theta / m^*$ and the bare detuning $\delta_0 = \omega_p - \omega_0$. Eq. \eqref{eq:continuity} is the continuity equation, while the gradient of Eq. \eqref{eq:euler} is analogous to the Euler equation.

When considering plane wave pump of the form $F(x) = \sqrt{I} e^{i k_p  x}$, we may seek a stationary ($\partial_t n = \partial_t \theta = 0)$ and homogeneous ($\partial_x n = \partial_x v =0$) solution of \eqref{eq:continuity} and \eqref{eq:euler}. By writing $\theta(x) = k_p x + \theta_0$, we conclude that
\begin{equation}
    0 = -\gamma n - 2\sqrt{nI}\sin \theta_0 ;
\end{equation}
\begin{equation}
    gn - \delta + \sqrt{\frac{I}{n}}\cos \theta_0 = 0.
\end{equation}
where we have introduced the detuning $\delta = \delta_0 - \frac{\hbar k_p^2}{2m} + V$. Then, we arrive at the solutions
\begin{equation}
    \left[ (gn - \delta)^2 + \frac{\gamma^2}{4} \right]n = I,
\end{equation}
\begin{equation}
    \theta_0 = \arctan \left[\frac{\gamma}{2(\delta - gn)} \right]
\end{equation}

\section{The truncated Wigner approximation}

The Truncated Wigner Method (TWM) allows one to efficiently simulate quantum expectation values and has been extensively employed in studies of polaritons \cite{carusotto_parametric_threshold_2005}. The master equation for this system in the Wigner representation may be approximated by a Fokker-Planck equation if the condition $\gamma \ll g / \Delta x$ is satisfied, where $\Delta x$ is the grid spacing used in the simulation. The corresponding Langevin equations in the Ito sense are given by
\begin{equation}
    \label{eq:ito}
    i d\psi(x, t) = A\{\psi\}(x)dt + \sqrt{\frac{\gamma}{2 \Delta x}} dW(x, t)
\end{equation}
Here, $A\{\psi\}(x)$ denotes the right hand side of Eq. \eqref{eq::ddgpe}. Furthermore, $dW(x, t)$ is a complex Wiener process, meaning that the increment
\begin{equation}
    \xi(x) = \int_{t_0}^{t_0 + \Delta t} dW(x, t) .
\end{equation}
is a complex Gaussian variable independent of $t_0$, satisfying $\expval{\xi(x)} = 0$ and $\expval{\xi(x)\xi^*(x^\prime)} = \delta_{x, x^\prime} \Delta t$. The initial conditions for this equation should be sampled from the Wigner distribution of the initial state. In the case of the vacuum, corresponding to an initially empty cavity, $\psi(x, 0)$ is also a complex Gaussian variable, such that $\expval{\psi(x,0)} = 0$ and $\expval{\psi(x,0)\psi^*(x^\prime,0)} = \delta_{x, x^\prime} / 2\Delta x$. Then, the averages over this stochastic process correspond to averages over symmetrically ordered products of the field operators:
\begin{equation}
    \expval{O_1 \cdots O_N}_W = \frac{1}{N!}\sum_P \expval{\hat{O}_{P(1)} \cdots \hat{O}_{P(N)}} .
\end{equation}
The sum runs over all permutations of the indices $P$.

Usually, one is interested in normally ordered observables, which have to be symmetrized in order to be calculated from the TWM. The first order correlation is given by
\begin{equation}
    G_1(x, x^\prime) = \expval{\hat{\psi}^\dagger(x) \hat{\psi}(x^\prime)} = \expval{\psi^*(x) \psi(x^\prime)}_W + \frac{\delta_{x, x^\prime}}{2\Delta x},
\end{equation}
while the second order correlation is given by
\begin{equation}
    \begin{aligned}
        G_2(x, x^\prime) &= \expval{\hat{\psi}^\dagger(x^\prime) \hat{\psi}^\dagger(x) \hat{\psi}(x) \hat{\psi}(x^\prime)} \\ &= \expval{\abs{\psi(x)}^2\abs{\psi(x^\prime)}^2}_W \\ & - \frac{1 + \delta_{x, x^\prime}}{2\Delta x} \expval{\abs{\psi(x)}^2 + \abs{\psi(x^\prime)}^2 - \frac{1}{2\Delta x}}_W
    \end{aligned}
\end{equation}

The normalized second order correlation
\begin{equation}
    g_2(x, x^\prime) = \frac{G_2(x, x^\prime)}{n(x) n(x^\prime)}
\end{equation}
with $n(x) = G_1(x, x)$ reveal some signatures of the Hawking effect.

For the windowed Fourier transform of the field
\begin{equation}
    \hat{\psi}_a(k, t) = \int_{0}^{L} dx\, \ee^{-\ci k  x} w_a(x) \hat{\psi}(x, t)
\end{equation}
the commutation relations are modified to
\begin{equation}
    \left[ \hat{\psi}_{a}(k), \hat{\psi}_{b}^{\dagger}(k^\prime) \right] = f_{ab}(k-k^\prime)
\end{equation}
where
\begin{equation}
    f_{ab}(k) = \int_{0}^{L} dx e^{-ik  x} w_a(x) w_b(x).
\end{equation}

This imposes modifications in the formulas for the momentum space correlation functions. The first order one is given by
\begin{equation}
    \begin{aligned}
        G_1^{ab}(k, k^\prime) &= \langle \hat{\psi}_{b}^{\dagger}(k^\prime)\hat{\psi}_{a}(k) \rangle \\ &= \langle \psi_{b}^{*}(k^\prime)\psi_{a}(k) \rangle_{W} - \frac{1}{2} f_{ab}(k-k^\prime),
    \end{aligned}
\end{equation}
with corresponding populations $n_a(k) = G_1^{aa}(k, k)$. In an analogous manner, the second order correlation is expressed as
\begin{equation}
    \begin{aligned}
       G_2(k, k^\prime) &= \langle \hat{\psi}_{u}^{\dagger}(k^\prime)\hat{\psi}_{d}^{\dagger}(k)\hat{\psi}_{d}(k)\hat{\psi}_{u}(k^\prime) \rangle \\ &= \langle \abs{\psi_u(k^\prime)}^2 \abs{\psi_d(k)}^2 \rangle_{W} \\
&- \frac{1}{2} \langle \abs{\psi_u(k^\prime)}^2 \rangle_{W} f_{dd}(0) \\
&- \frac{1}{2} \langle \abs{\psi_d(k)}^2 \rangle_{W} f_{uu}(0) \\
&- \Re \left[\langle \psi_{u}^{*}(k')\psi_{d}(k) \rangle_W f_{ud}(k-k^\prime) \right] \\
&+ \frac{f_{uu}(0)f_{dd}(0)}{4} \\
&+ \frac{\abs{f_{ud}(k-k^\prime)}^2}{4} .
\end{aligned}
\end{equation}
Finally, the normalized second order correlation is given by $g_2(k, k^\prime) = G_2(k, k^\prime)/ n_u(k^\prime)n_d(k)$.

\section{Considerations on window placement}

\begin{figure}
    \centering
    \includegraphics[width=\linewidth]{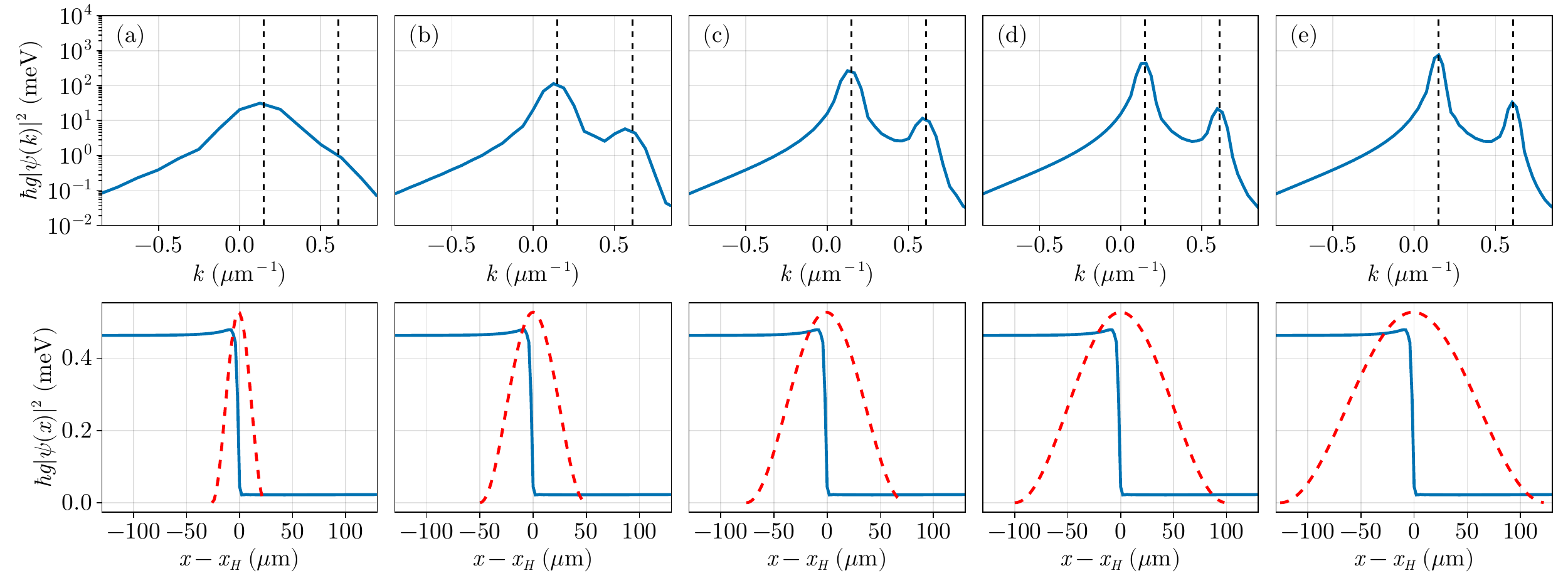}
    \caption{Bottom row: different windows shown over the mean field position density. Top row: corresponding mean field momentum density. The windows are represented by dashed red lines. The black vertical lines correspond to the pump wavenumber $k_{up}$ and $k_{down}$.}
    \label{fig:momentum_density_width}
\end{figure}

\begin{figure}
    \centering
    \includegraphics[width=\linewidth]{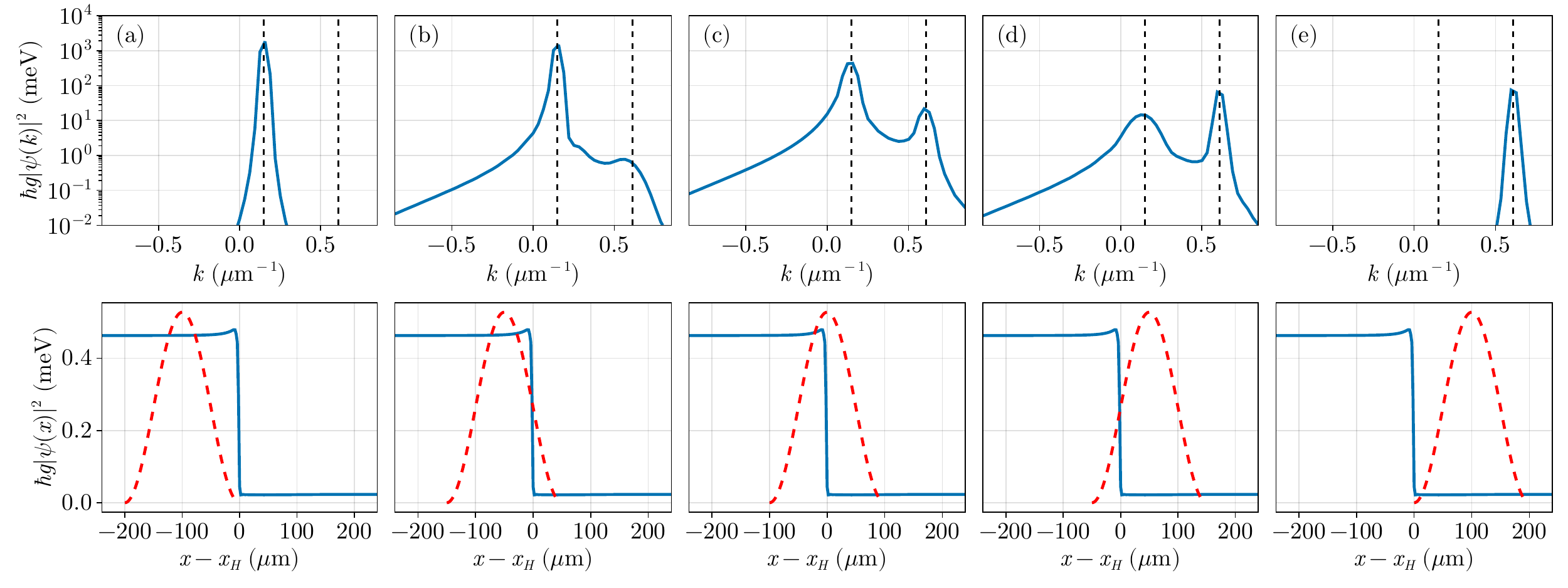}
    \caption{Bottom row: different windows shown over the mean field position density. Top row: corresponding mean field momentum density. The windows are represented by dashed red lines. The black vertical lines correspond to the pump wavenumber $k_{up}$ and $k_{down}$.}
    \label{fig:momentum_density_center}
\end{figure}

\begin{figure}
    \centering
    \includegraphics[width=\linewidth]{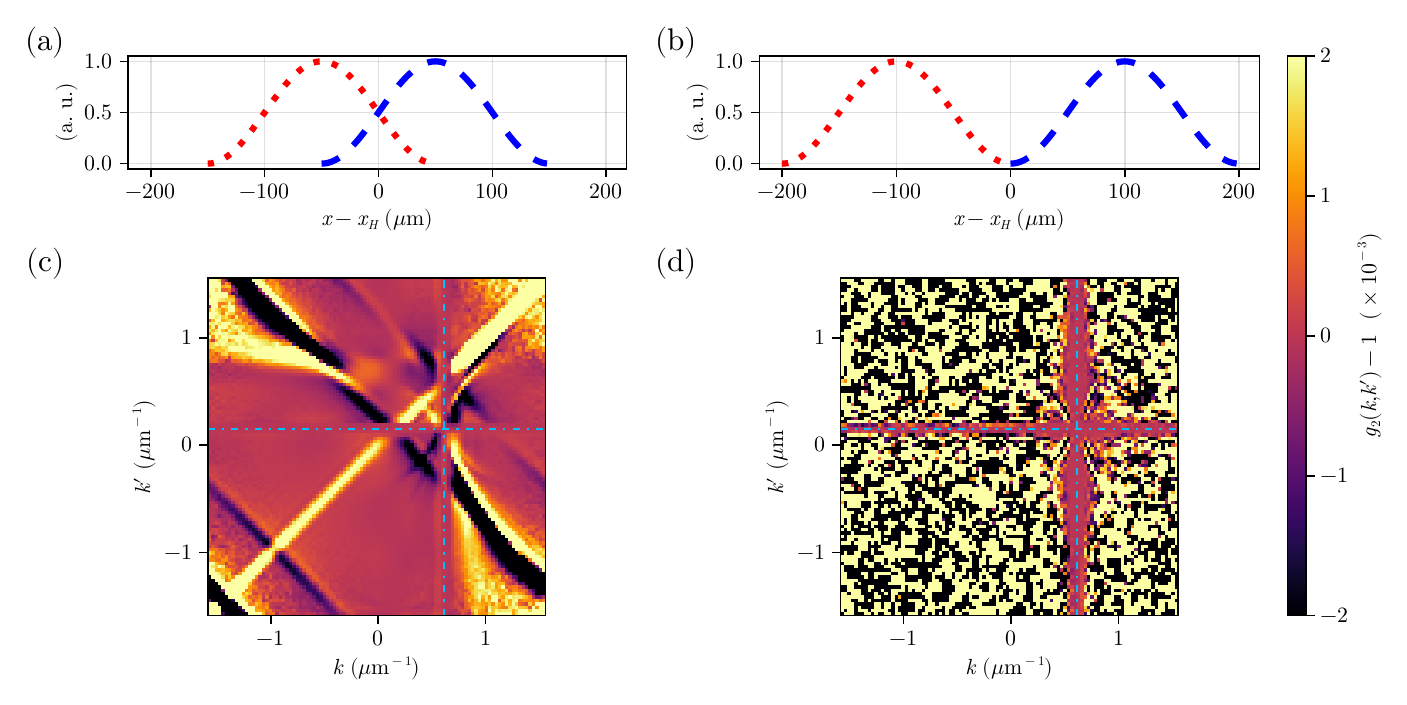}
    \caption{(a) and (b) correspond to the pairs of windows that generate the momentum density correlations in (c) and (d), respectively. In this simulation, $1.7 \times 10^8$ realizations were performed.}
    \label{fig:more_windows}
\end{figure}

The use of a windowed Fourier transform serves many purposes in our context. From the physical point of view, by placing two windows on opposite sides of the horizon, we are able to more clearly isolate the Hawking radiation from other signals that appear in $g_2$, as the strong autocorrelations, for example. Furthermore, the methods usually employed to solve the equations of motion assume periodic boundary conditions. Then, one needs to modify both the pump and the losses close to the edges in order to stably support the fluid at the desired working points, as shown in Fig. \ref{fig:potential_losses_pump}. These artifacts lead to the appearance of spurious signals in the momentum space correlation if one also includes the boundaries in the calculation.

However, proper use of window functions requires careful consideration of both its size and placement. The length $L_w$ of a window defines the resolution $\Delta k_w = 2\pi / L_w$ in momentum space that can be resolved. This is a reflex of the uncertainty relation between Fourier pairs. Therefore, a localized window gives poor momentum resolution. To illustrate this point, we show, in Fig. \ref{fig:momentum_density_width}, the momentum density $\abs{\psi(k)}^2$ for different window lengths, all centered at the horizon. As both the upstream and downstream regions fall under the window, we should see two peaks corresponding to $k_{up}$ and $k_{down}$. Although they are clearly visible in \ref{fig:momentum_density_width}(e), when the window is large, they become less pronounced until they merge completely in \ref{fig:momentum_density_width}(a). This spread of the peaks is a manifestation of spectral leakage.

In order to eliminate autocorrelation signals in $g_2$, one would like to place windows completely contained on different sides of the horizon, as shown in Fig. \ref{fig:momentum_density_center}(a,e). However, this placement only leaves us with the sharp population peaks corresponding to the pump. This is due to the dissipative nature of the polaritons, which causes the other momentum components created at the horizon to leave the cavity before reaching the maximum of the window. In this case, we would not be able to observe the Hawking effect, as there would be no population in the required modes. As discussed above, one also cannot use small windows, as that would cause a severe spectral leakage. The best compromise, as employed in the main text, is to place the windows only partially in each region, as in Fig. \ref{fig:momentum_density_center}(b,d). This ensures a sufficient momentum resolution, a significant damping of the autocorrelations, while also allowing for the Hawking radiation to reach the maximum of the window before leaving the cavity.

To further illustrate this point, we show, in Fig. \ref{fig:more_windows}, the value of $g_2$ for two more different pairs of windows. One can see that, when the windows are fully contained within each region, as shown in (b), the corresponding momentum correlations in (d) display only noise, which is consistent with an extremely low occupation number of the corresponding wavevectors.

\section{Stabilization of the fluid in the numerical simulation}

\begin{figure}
    \centering
    \includegraphics[width=0.6\linewidth]{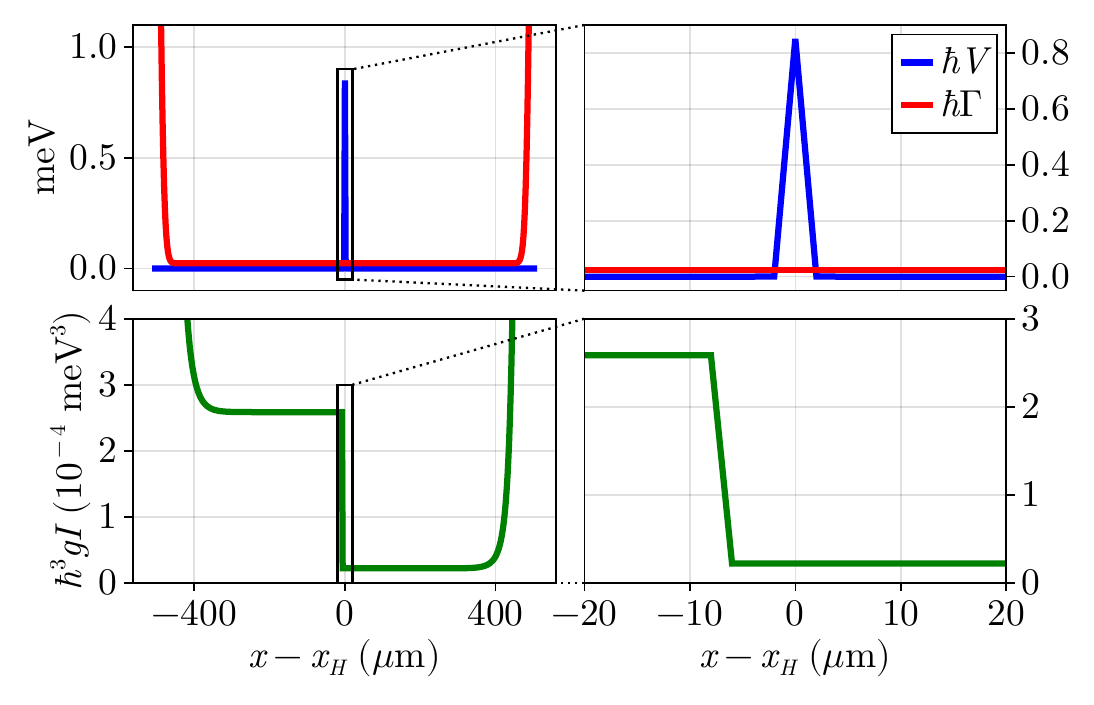}
    \caption{Top row: potential and losses; bottom row: intensity profile.}
    \label{fig:potential_losses_pump}
\end{figure}

Due to the fact that we are working very close to the critical point of the bistability cycle, it is hard, both numerically and experimentally, to stabilize the polariton fluid in the desired working point. This is further complicated in the numerical case, as the utilized algorithm assumes periodic boundary conditions. In this section we describe the prescription used to achieve a flat and stable configuration of the fluid in both the upstream and downstream regions.

First, to avoid undesired effects arising from the periodic boundary conditions, we use a modified spatially dependent loss
\begin{equation}
    \Gamma(x) = \gamma + \gamma_{damp} [ G(x; 0,w_{damp}) + G(x; L, w_{damp}) ]
\end{equation}
which rises significantly close to the edges of the simulated cavity. In the above equation, $G$ denotes a Gaussian function: $G(x; x_0, w) = e^{-(x-x_0)^2/w^2}$.

The pump is also modified to rise in the edges, up to a value $F_M$:
\begin{equation}
    F(x,t) = \begin{cases}
        [F_{up} + (F_{M} - F_{up})\sech(x/w_p)]e^{ik_{up}x-\omega_p t} h(t), & x< x_s \\
        [F_{down} + (F_{M} - F_{down})\sech((x-L)/w_p)]e^{ik_{down}x-\omega_p t} h(t), & x> x_s
    \end{cases}
\end{equation}
Here, $x_s$ is the position of the step of the pump. Furthermore,
\begin{equation}
    h(t) = 1 + Ae^{-t/\tau}
\end{equation}
is a term representing an extra amplitude $A$ at the start of the pumping scheme in order to rise the fluid to the top of the bistability curve. This term decays after a transient time $\tau$.

We also included a potential $V(x) = V_0 G(x; x_d, w_V)$ localized around the position $x_d$.

All of these terms are illustrated in Fig. \ref{fig:potential_losses_pump}

\section{Numerical parameters}

In table \ref{tab:parameters}, we display the numerical values of all the relevant parameters that were used in the simulations presented in this work.

\begin{table}[]
    \centering
    \begin{tabular}{ccc} \toprule
        Parameter name & Parameter symbol & Parameter value \\ \midrule
        Cavity length & $L$ & 1024 $\mu$m \\
        Number of points in the grid & $N$ & 512 \\
        Grid spacing & $\Delta x$ & 2 $\mu$m \\
        Time step of the simulation & $\Delta t$ & 0.2 ps \\
        Simulation duration & $T$ & 400 ps \\
        Number of realizations & $N_{\mathrm{sim}}$ & $5.5 \times 10^8$ \\
        Planck's constant & $\hbar$ & 0.6582 meV $\cdot$ ps \\
        Loss & $\gamma$ & 0.047 meV$/\hbar$ \\
        Polariton mass & $m^*$ & 1/6 meV $\cdot$ ps$^{2}$ $\cdot$ $\mu$m$^{-2}$ \footnote{This is roughly $3 \times 10^{-5}$ the mass of the electron} \\
        Detuning & $\delta(0)$ & 0.49 meV$/ \hbar$ \\
        Nonlinear constant & $g$ & $3 \times 10^{-4}$ meV $\cdot$ $\mu$m$/ \hbar$ \\
        Extra damping at the edges & $\gamma_\mathrm{damp}$ & 4.5 meV$/\hbar$ \\
        Width of the damping at the edges & $w_\mathrm{damp}$ & 20$\mu$m \\
        Upstream pump momentum & $k_\mathrm{up}$ & 0.148 $\mu$m$^{-1}$ \\
        Downstream pump momentum & $k_\mathrm{down}$ & 0.614 $\mu$m$^{-1}$ \\
        Upstream pump amplitude & $F_\mathrm{up}$ & 1.410 ps$^{-1}$ $\cdot$ $\mu$m$^{-1/2}$ \\
        Downstream pump amplitude & $F_\mathrm{down}$ & 0.410 ps$^{-1}$ $\cdot$ $\mu$m$^{-1/2}$\\
        Maximum pump amplitude & $F_\mathrm{M}$ & 20 ps$^{-1}$ $\cdot$ $\mu$m$^{-1/2}$\\
        Width of the pump at the edge & $w_\mathrm{p}$ & 20$\mu$m\\
        Position of the potential step & $x_\mathrm{s}$ & 505 $\mu$m \\
        Extra amplitude factor & $A$ & 6 \\
        Transient time & $\tau$ & 50ps \\
        Position of the deffect & $x_\mathrm{d}$ & 512 $\mu$m \\
        Potential amplitude & $V_0$ & 0.85 meV$/\hbar$ \\
        Width of the potential & $w_\mathrm{V}$ & 0.75 $\mu$m
    \end{tabular}
    \caption{Numerical values of the parameters relevant to the simulation.}
    \label{tab:parameters}
\end{table}

\section{From momentum–space \texorpdfstring{$g^{(2)}$}{g2} to photocurrent correlations}
\label{sec:detection}
Let the total optical power collected from window \(a\in\{u,d\}\) be \(P_a\). The mean photo‑electron rate on detector \(a\) is
\begin{equation}
R_a=\eta_a\,\mathcal{T}_a\,\frac{P_a}{\hbar\omega_0},
\end{equation}
with overall quantum efficiency \(\eta_a\) and optical throughput \(\mathcal{T}_a\).
The normally ordered, time‑resolved cross‑correlation of the photocurrents is
\begin{equation}
C_{ud}(\tau)\equiv \langle :\delta i_u(t)\,\delta i_d(t+\tau):\rangle
= e^2 R_u R_d\,\big[g^{(2)}_{ud}(\tau)-1\big],
\end{equation}
where \(g^{(2)}_{ud}(\tau)\) is the \emph{time‑domain} second‑order correlation integrated over the momentum windows defined by \(w_{u,d}\).

In continuous‑wave operation with sampling bin \(\Delta t\) comparable to the correlation time \(\tau_c\) (set by the polariton lifetime and detection bandwidth), the expected number of coincidences accumulated over a total integration time \(T\) is
\begin{equation}
N_{\rm acc}= R_u R_d\,\Delta t\,T,\qquad
N_{\rm true}= R_{\rm pair}\,T,
\end{equation}
and the measured zero‑delay correlation is
\begin{equation}
g^{(2)}_{ud}(0) \simeq 1+\frac{N_{\rm true}}{N_{\rm acc}}
= 1+ \frac{R_{\rm pair}}{R_u R_d\,\Delta t}.
\label{eq:g2pairs}
\end{equation}
Comparing with the definition used in the main text at equal times,
\begin{equation}
\delta g^{(2)} \equiv g^{(2)}_{ud}(0)-1,
\end{equation}
we identify the pair rate into the two detected momentum windows as
\begin{equation}
R_{\rm pair} = \delta g^{(2)}\, R_u R_d\,\tau_c,
\qquad
(\Delta t\sim\tau_c).
\label{eq:Rpair}
\end{equation}
Equation~\eqref{eq:Rpair} allows to interpret the numerically obtained \(\delta g^{(2)}\) as an excess coincidence probability per correlation time (not a fractional increase of the singles rate).
Note that the dependence is quadratic in the singles and linear in the correlation time.

When \(\delta g^{(2)}\ll 1\) (our case), accidental coincidences dominate the variance, giving
\begin{equation}
\mathrm{SNR} \simeq \frac{N_{\rm true}}{\sqrt{N_{\rm acc}}}
= \delta g^{(2)}\sqrt{R_u R_d\,\tau_c\,T}.
\end{equation}
To reach a significance \(s\) one needs
\begin{equation}
T \simeq \frac{s^2}{\big[\delta g^{(2)}\big]^2\,R_u R_d\,\tau_c}\,.
\label{eq:time}
\end{equation}
For illustration: if each arm carries \(R_u=R_d=10^8~\mathrm{s}^{-1}\) (easily obtained after attenuation from \(P_a\sim 0.1~\mu\mathrm{W}\) at near‑IR frequencies with Si or InGaAs photodiodes operated linearly), \(\tau_c\sim 10~\mathrm{ps}\), and \(\delta g^{(2)}\sim 10^{-3}\), then \(R_u R_d \tau_c\simeq 10^5\) and a \(5\sigma\) detection requires \(T\simeq 25/(10^{-6}\times 10^5)\approx 250~\mathrm{s}\), that is, about four minutes.

Balanced detection is used to suppress classical common‑mode noise from the pump. Practically, each momentum window is split 50:50; one output goes to the ``signal'' photodiode and the other to a reference diode.
The two currents of each pair are subtracted (with appropriate gain balancing), producing shot‑noise‑limited residuals \(\tilde i_u,\tilde i_d\).
The Hawking signal is obtained from the cross‑correlation \(\langle \tilde i_u(t)\tilde i_d(t+\tau)\rangle\).
Because the subtraction is local to each arm, the mean singles rates in Eq.~\eqref{eq:time} must be understood as the post‑split rates; the improvement comes from removing classical intensity noise without affecting the normally ordered cross‑correlation, which is preserved by passive linear optics.

Alternatively, one may estimate the zero‑frequency cross‑spectrum \(S_{ud}(0)=2e^2 R_u R_d\!\int\!d\tau\,[g^{(2)}_{ud}(\tau)-1]\), using RF spectrum analyzers with bandwidth \(B\).
The integrated SNR then scales as \(\propto \delta g^{(2)}\sqrt{R_u R_d\,\tau_c\,T}\), identical to Eq.~\eqref{eq:time}.
Fast time resolution is unnecessary, provided that the electronics bandwidth exceeds \(1/\tau_c\).

%